\documentclass{emulateapj}
\usepackage{graphicx}
\begin{document}

\def\etal{et al.\ \rm}
\def\ba{\begin{eqnarray}}
\def\ea{\end{eqnarray}}
\def\etal{et al.\ \rm}

\title{Metal accretion onto white dwarfs caused by Poynting-Robertson 
drag on their debris disks.}

\author{Roman R. Rafikov\altaffilmark{1,2}}
\altaffiltext{1}{Department of Astrophysical Sciences, 
Princeton University, Ivy Lane, Princeton, NJ 08540; 
rrr@astro.princeton.edu}
\altaffiltext{2}{Sloan Fellow}


\begin{abstract}
Recent discoveries of compact (sizes $\lesssim$ R$_\odot$) 
debris disks around more than a dozen of metal-rich white dwarfs (WDs) 
suggest that pollution of these stars with metals may 
be caused by accretion of high-Z material from the disk. 
But the mechanism responsible for efficient transfer of 
mass from a particulate disk to the WD atmosphere 
has not yet been identified. Here we demonstrate that
radiation of the WD can effectively drive accretion of matter 
through the disk towards the sublimation radius (located at several tens of 
WD radii), where particles evaporate, feeding a disk of metal 
gas accreting onto the WD. We show that, contrary to some previous claims,  
Poynting-Robertson (PR) drag on the debris disk is effective at providing 
metal accretion rate $\dot M_{PR}\sim 10^8$ g s$^{-1}$ and
higher, scaling quadratically with WD effective temperature. 
We compare our results with observations and show that, as 
expected, no WD hosting a particulate debris disk shows evidence 
of metal accretion rate below that produced by the PR drag. 
Existence of WDs accreting metals at 
rates significantly higher than $\dot M_{PR}$ suggests that 
another mechanism in addition to PR drag drives 
accretion of high-Z elements in these systems.  
\end{abstract}



\section{Introduction.}  
\label{sect:intro}


Recent infrared observations with {\it Spitzer} and ground-based 
facilities (Zuckerman \& Becklin 1987; Graham \etal 1990; 
Farihi \etal 2010) revealed near-infrared excesses 
around more than a dozen of metal-rich white dwarfs (WDs).
This emission was interpreted as reprocessing  
of the WD radiation by refractory material residing in an 
extended, optically thick and geometrically thin disk (Jura 2003; 
Jura \etal 2007), similar to the rings of Saturn
(Cuzzi \etal 2010). 

It was hypothesized by Jura (2003) that such compact disks of high-Z
material may be naturally produced 
by tidal disruption of asteroid-like bodies entering the 
Roche radius of the WD. This idea naturally explains the 
well defined outer radii $R_{out}\sim 1$ R$_\odot$
of disks since the Roche radius $R_R$ for tidal disruption 
of a self-gravitating object of normal density 
($\rho\sim 1$ g sm$^{-3}$) by $M_\star\sim$ M$_\odot$ central 
mass is $R_R\sim (M_\star/\rho)^{1/3}\approx 1$ R$_\odot$ 
(Jura 2003). 

Moreover, Jura (2003) went on to suggest that 
the high-Z material contained in compact disks is 
responsible for the observed metal enrichment of 
a significant fraction of WDs. Accretion of this 
{\it circumstellar} material at high enough rate 
$\dot M_Z$ onto WD could maintain a 
non-zero abundance of metals in the WD atmosphere 
against rapid gravitational settling.
This scenario thus provides a promising alternative to the 
previously widely discussed {\it interstellar} accretion model of 
WD metal pollution (Dupuis \etal 1993), which is known
to have serious problems. 

Observed abundances of heavy elements in WD
atmospheres and theoretical calculations of their gravitational 
settling imply typical metal accretion rates 
$\dot M_Z\sim 10^6-10^{10}$ g s$^{-1}$ 
(Farihi \etal 2009, 2010). An evolving disk of debris must 
be able to supply such high $\dot M_Z$ to the WD.
However, the question of how high-Z 
elements get transported to the WD atmosphere from the ring of 
solid particles, which does not extend all the way to the WD 
surface, has not yet been answered. 

In principle, a dense ring of particles should evolve 
simply because of the angular momentum transfer due to
inter-particle collisions, in full analogy with 
the rings of Saturn. However, the evolution time 
scale of Saturn's rings due to this process is too long, 
$\sim 10^9$ yr (Salmon \etal 2010), and resultant values 
of $\dot M_Z$ are negligible.

Another natural mechanism driving debris towards the WD is 
due to stellar radiation interacting with the disk and giving rise 
to the Pointing-Robertson (PR) drag (Burns \etal 1979). 
Previously, 
Farihi \etal (2010) claimed that PR drag cannot provide
$\dot M_Z$ higher than $10^3-10^4$ g s$^{-1}$ and dismissed 
this process as irrelevant. 
The goal of this paper is to critically re-examine the effect of 
radiative forces on the debris disk evolution, and to show in 
particular that PR drag can give rise to $\dot M_Z$ inferred from 
observations.


\section{Mass accretion due to radiative forces.}
\label{sect:theory}


We envisage the following conceptual picture of the circum-WD 
environment. A dense disk (or ring) of particles lies inside the 
Roche radius $R_R$ of the WD and evolves under the action of external 
agents, e.g. radiation forces. Particles migrate through the disk 
towards the sublimation radius $R_s$, where their equilibrium 
temperature equals the sublimation temperature $T_s$:
\ba
R_s=\frac{R_\star}{2}
\left(\frac{T_\star}{T_s}\right)^{2}\approx 22~R_\star
T_{\star,4}^2
\left(\frac{1500\mbox{K}}{T_s}\right)^2,
\label{eq:R_S}
\ea
where $R_\star$ is the WD radius, $T_s\approx 1500$ K for 
silicate grains, and $T_{\star,4}\equiv T_\star/10^4$ K is 
the normalized stellar temperature $T_\star$.  
Taking $R_\star\approx 0.01R_\odot$ typical for massive 
($M_\star\gtrsim 0.6$ M$_\odot$) WDs 
(Ehrenreich \etal 2011), one finds $R_s\approx 0.2$ R$_\odot$, 
in agreement with observationally inferred inner radii of
compact debris disks (Jura \etal 2007, 2009a). Thus, for cool 
WDs $R_{out}/R_{in}$ is several.

Particles sublimate at $R_s$ feeding a disk 
of metallic gas, which we assume to be transparent 
to stellar radiation (gaseous component has been detected in 
several WDs with debris disks, see G\"ansicke etal 2006). 
Gaseous disk viscously evolves, extending all the way to the WD 
surface and providing means of metal delivery  
from $R_s$ to the star. 
Assuming conventional $\alpha$-parametrization of viscosity 
(Shakura \& Sunyaev 1973) $\nu=\alpha c_s^2/\Omega_K$, 
the characteristic viscous time in the disk of metallic gas is 
\ba
t_\nu\sim \frac{R_s^2}{\nu}\approx 10~\mbox{yr}~
\frac{\mu_{28}}{\alpha}\frac{1500~\mbox{K}}{T}
\left(M_{\star,1}
\frac{R_s}{0.2~\mbox{R}_\odot}\right)^{1/2},
\label{eq:t_nu}
\ea
where $M_{\star,1}\equiv M_\star/M_\odot$, 
and $\mu_{28}$ is the mean molecular weight of the metallic 
gas normalized by $28m_p$ (value of $\mu$ for pure Si).
This timescale is short enough for $\dot M_Z$ to be determined 
predominantly by evolution of the disk of solids. 

We take the disk of particles to be 
geometrically thin and characterize
it at each point by optical depth $\tau$ defined as
\ba
\tau\equiv \frac{3}{4}\frac{\Sigma_d}{\rho a},
\label{eq:tau}
\ea
where $\Sigma_d$ is the surface mass density of debris disk, and 
$\rho$ and $a$ are the bulk density and characteristic size of
the constituent particles (we leave $a$ unspecified for now). 
Observations suggest (Jura 2003; Jura \etal 2007) that 
$\tau\gtrsim 1$ but we leave $\tau$ arbitrary. 

\begin{figure}
\plotone{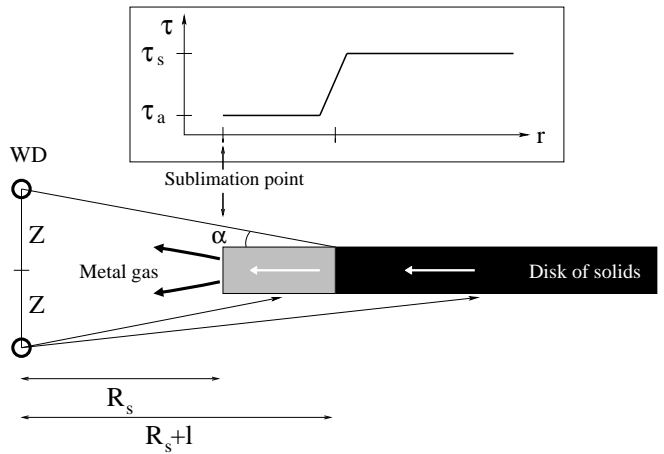}
\caption{
``Lamp-post'' model of debris disk illumination by the WD at small 
incidence angle $\alpha$. WD emission is represented by two point 
sources at height $\pm Z$ to account for the finite size of the WD.
An inner edge of the disk of solids (black) near the sublimation 
point at $R_s$ is shown, including an annulus of directly exposed 
solid material (gray). White arrows indicate accretion of solids,
black arrows indicate accretion of metal gas inside of $R_s$. 
Disk thickness $h\ll R_\star$ is not shown to scale.
Inset on top schematically shows the run of the disk optical depth
with radius. See text for details.
\label{fig:f1}}
\end{figure}

Particles in a narrow annulus just outside the sublimation 
radius are exposed to direct starlight at normal incidence
and are heated to sublimation temperature.
Farihi \etal (2010) studied the effect of PR 
drag on particles in this region. Here we first 
explore the effect of radiative forces on the rest of the 
disk, which has only its surface illuminated by the central 
star at grazing incidence. To account for the finite size of 
the WD we adopt
a ``lamp-post'' model of stellar illumination, in which 
radiation is emitted by two point sources of luminosity $L_\star/2$
each located at height 
$Z=(4/3\pi)R_\star$ above and below the disk plane, 
see Figure \ref{fig:f1}. Then 
\ba
\alpha\approx \tan\alpha=\frac{4}{3\pi}\frac{R_\star}{r} 
\label{eq:alpha}
\ea
is the incidence angle of radiation on the disk surface at 
distance $r$ from the star, and $\alpha\ll 1$ since 
$r\ge R_s\gg R_\star$. The one-sided incident energy 
flux per unit surface area of the disk is 
$F_r=(L_\star/6\pi^2)(R_\star/r^3)$, in agreement 
with Friedjung (1985).

Radiation falling on the disk gives rise to two forces 
that can change angular momentum of the disk: PR drag
and the Yarkovsky force. The latter owes
its existence to thermal inertia of spinning objects that causes
re-emission of absorbed energy in the direction different  
from that of incoming radiation (Bottke \etal 2006).
Azimuthal components of these 
forces per unit area of the disk can be represented as
\ba
f_\varphi=
\alpha\frac{L_\star \phi_r}{4\pi r^2 c}\psi,
\label{eq:f_PR}
\ea
where $c$ is the speed of light, and $\psi$ is the factor that 
characterizes the phase lag of radiative momentum deposition 
in disk particles and is specified later in 
\S\ref{subsect:PR} and \S\ref{subsect:Yar} for PR drag 
and Yarkovsky force respectively. Factor $\phi_r$ characterizes 
the efficiency of radiative momentum absorption by the disk surface.
In the geometrical optics limits (particle size $a$ much larger
than characteristic wave length of stellar radiation) 
\ba
\phi_r\approx 1-e^{-\tau_\parallel},~~~\tau_\parallel\equiv 
\alpha^{-1}\tau,
\label{eq:Q}
\ea
where $\tau_\parallel$ is the optical depth encountered 
by incident photons as they traverse the full disk thickness.
Optically thick debris disks have $\phi_r= 1$.

Azimuthal force causes radial drift of disk material at speed 
$v_r=2f_\varphi/(\Omega_K\Sigma_d)$, where $\Omega_K$ is the Keplerian 
angular frequency. This gives rise to mass transport at the rate 
\ba
\dot M_Z(r)=2\pi r v_r\Sigma_d=\alpha\psi
\frac{L_\star\phi_r}{\Omega_K r c}.
\label{eq:Mdot_r}
\ea
Note that for radiative forces both $f_\varphi$ and $\dot M_Z$ 
are independent of $\Sigma_d$.


\subsection{Poynting-Robertson drag.}
\label{subsect:PR}

In the case of PR drag factor $\psi$ is (Burns \etal 1979)
\ba
\psi_{PR}=\frac{\Omega_K r}{c}\approx 3.3\times 10^{-3}
\left(M_{\star,1}\frac{0.2\mbox{R}_\odot}{r}
\right)^{1/2}
\label{eq:psiPR}
\ea
leading to
\ba
\dot M_{PR}(r)&=&\frac{4 \phi_r}{3\pi}
\frac{R_\star}{r}
\frac{L_\star}{c^2}
\label{eq:Mdot_PR}\\
&\approx & 10^8\mbox{g s}^{-1}\phi_r
\frac{L_\star}{10^{-3}L_\odot}\frac{20}{r/R_\star},
\nonumber
\ea
where we used equations (\ref{eq:alpha}) and (\ref{eq:Mdot_r}).

Numerical estimate in equation (\ref{eq:Mdot_PR}) does not 
agree with the calculation of Farihi \etal (2010) and we 
demonstrate the origin of this discrepancy by following
these authors and considering the inner edge of the disk 
where particles sublimate (at $r=R_s$). At the edge there 
is an annulus of the disk directly exposed to starlight 
(i.e. $\tau_{\parallel,a}\lesssim 1$, see Figure 
\ref{fig:f1}) with radial width 
\ba
l\sim \tau_a^{-1}h,
\label{eq:l}
\ea
which follows from the relationship 
$\tau_a/h\sim\tau_{\parallel,a}/l$ (here $h\ll R_\star$ 
is the disk thickness, $\tau_a$ and 
$\tau_{\parallel,a}\lesssim 1$ are the vertical and 
horizontal optical depths of the annulus).
Particles inside this annulus experience PR drag as individual 
objects, with little shadowing from nearby particles. 
The characteristic decay time of their semimajor axes is
(Burns \etal 1979)
\ba
t_{PR}=\frac{4\pi}{3}\frac{\rho a R_s^2 c^2}{L}.
\label{eq:t_PR}
\ea

However, it only takes the fully exposed particles time
$t_{cross}\sim t_{PR}(l/R_s)$
to cross the annulus, reach the sublimation point, and evaporate, 
at which point they stop shadowing objects behind them. 
Dividing the mass $2\pi R_s l\Sigma_d$ contained within the annulus 
by $t_{cross}$ we derive the mass accretion rate inside the annulus:
\ba
\dot M_{a}\sim \frac{2\pi R_s l\Sigma_d}{t_{PR}}\frac{R_s}{l}
\sim \tau_a\frac{L}{c^2},
\label{eq:dotM_a}
\ea
where we used definition (\ref{eq:tau}). Mass conservation
requires that $\dot M_{a}=\dot M_{PR,s}\equiv\dot M_{PR}(R_s)$ 
so that the optical depth inside the annulus is 
$\tau_a\sim \alpha_s\phi_r\ll 1$ 
($\alpha_s\equiv\alpha(R_s)$) and from equation (\ref{eq:l}) 
the width of the annulus is $h/(\alpha_s\phi_r)$. Thus, as 
long as the disk outside the annulus is not transparent to 
incident stellar radiation (i.e. $\tau_\parallel\gtrsim 1$
or $\tau\gtrsim\alpha$) $\tau$ drops from its value
$\tau_s$ outside of $R_s+l$ to $\tau_a$ inside. This is because fully
exposed particles in the annulus experience stronger PR drag and
migrate towards the WD faster than particles outside the annulus. 
The transition occurs at the distance $l$ outside the disk edge, 
where $\tau_{\parallel,a}\sim 1$, see Figure \ref{fig:f1}. 

From observational point of view, the most interesting characteristic
of the disk is 
the value of $\dot M_{PR}$ at sublimation point $R_s$, since this 
is the rate at which disk of solids feeds gaseous 
disk, from 
which mass is transported onto the WD atmosphere by viscous torques.
Evaluating expression (\ref{eq:Mdot_PR}) at $r=R_s$ using 
equation (\ref{eq:R_S}) we find
\ba
\dot M_{PR,s}&=&\frac{32\phi_r}{3}\sigma
\left(\frac{R_\star T_\star T_s}{c}\right)^2
\label{eq:massflux}\\
&\approx  &
7\times 10^7\mbox{g s}^{-1}\phi_r\left(R_{\star,-2}
T_{\star,4}\frac{T_s}{1500\mbox{K}}\right)^2,
\nonumber
\ea
where $R_{\star,-2}\equiv R_\star/10^{-2}$ R$_\odot$; then 
$\dot M_{PR}(r)=\dot M_{PR,s}(R_s/r)$.
Note that $\dot M_{PR,s}$ depends on $T_\star$ only 
quadratically.

Previously,  Graham \etal (1990) obtained an empirical 
estimate of $\dot M_{PR}$ by assuming accretion to be 
driven by PR drag and dividing the observationally 
inferred disk mass (surface area multiplied by $\rho a$, 
taking disk to be a monolayer of particles) by $t_{PR}$, 
see equation (\ref{eq:t_PR}). Even though their 
numerical estimate of $\dot M_{PR}\approx 1.8\times 10^8$ g
s$^{-1}$ is not very different from reality this method of
deriving $\dot M_{PR}$ is not well motivated.
Later, Farihi \etal (2010) found a value of 
$\dot M_{PR}\lesssim 10^4$ g s$^{-1}$ much lower than  
$\dot M_{PR,s}$ predicted by equation (\ref{eq:massflux}) 
because they used $t_{PR}$ instead of $t_{cross}\ll t_{PR}$ 
in computing $\dot M_{a}$.


\subsection{Yarkovsky force.}
\label{subsect:Yar}

If circum-WD debris 
disks behave similarly to dense planetary rings one expects
collisions in the disk to align particle spins predominantly 
parallel or anti-parallel to the disk normal, with non-zero 
and {\it positive} average spin (prograde mean rotation) and 
significant random spin component (Salo 1987; Ohtsuki 
\& Toyama 2005). The characteristic spin frequency is 
$\omega\sim \Omega_K$. Thermal inertia of spinning particles 
causes re-emission of absorbed stellar energy in the 
direction different from the radial. This gives rise to azimuthal 
Yarkovsky force, which causes {\it outward} orbital migration 
of disk particles spinning in prograde sense (Bottke \etal 2006).
We now check whether this force can affect the PR-driven 
inward accretion of solids by estimating its typical magnitude.

Yarkovsky phase lag factor $\psi_Y$ depends on two 
dimensionless parameters. One of them $R^\prime=a/l_\nu$ is 
the ratio of particle size $a$ to the penetration depth of 
the thermal wave $l_\nu=\sqrt{K/\rho C_p\omega}$, where
$K$ and $C_p$ are the thermal conductivity and specific heat
of particle material. Another is the thermal parameter (roughly the 
ratio of thermal time at depth $l_\nu$ to object's spin period)
$\Theta=\sqrt{K\rho C_p\omega}/\sigma T^3$, where $T$ is the 
surface temperature of the body. The actual dependence of
$\psi_Y$ on $R^\prime$ and $\Theta$ has been calculated 
in Vokrouhlick\'y (1998, 1999) and these references should
be consulted for details. For our current purposes it suffices 
to note that phase lag factor takes on a maximum value 
$\psi_{Y,max}\sim 0.1$ when $R^\prime\gtrsim 1$ and 
$\Theta\sim 1$, and is smaller for other values 
of $R^\prime$ and $\Theta$. 

To evaluate $l_\nu$ and $\Theta$ we assume that disk particles 
have composition typical for terrestrial silicate minerals 
such as olivine, in agreement with elemental abundances of 
metal-rich WD atmospheres (Zuckerman \etal 2007; Klein \etal 2010). Then
$K\approx 1.5\times 10^5$ erg s$^{-1}$ cm$^{-1}$ K$^{-1}$ and 
$C_p\approx 10^7$ erg g$^{-1}$ K$^{-1}$ at $T\gtrsim 10^3$ K 
(Roy \etal 1993). Taking $\omega=\Omega_K$ we can write
\ba
l_\nu=l_{\nu,s}\left(\frac{r}{R_s}\right)^{3/4},~~~
\Theta=\Theta_s\left(\frac{r}{R_s}\right)^{3/2},
\label{eq:pars}
\ea
where $l_{\nu,s}$ and $\Theta_s$ are the values of $l_\nu$ and
$\Theta$ at sublimation radius $r=R_s$:
\ba
l_{\nu,s}&=& \left(\frac{K}{\rho C_p}\right)^{1/2}
\left(\frac{R_\star^3}{8GM_\star}\right)^{1/4}
\left(\frac{T_\star}{T_s}\right)^{3/2}
\label{eq:lnus}\\
&\approx & 1~\mbox{cm}~
\frac{R_{\star,-2}^{3/4}T_{\star,4}^{3/2}}
{M_{\star,1}^{1/4}}
\left(\frac{1500\mbox{K}}{T_s}\right)^{3/2},
\nonumber\\
\Theta_s&=& \left(\frac{3\pi}{8}\right)^{3/4}
\frac{\left(K\rho C_p\right)^{1/2}}{\sigma T_s^3}
\left(\frac{GM_\star}{R_\star^3}\right)^{1/4}
\label{eq:Thetas}\\
&\approx & 10~
\frac{M_{\star,1}^{1/4}}{R_{\star,-2}^{3/4}}
\left(\frac{1500\mbox{K}}{T_s}\right)^{3}.
\nonumber
\ea
Particle size $a$ needed for calculation of $R^\prime$ is not 
constrained by observations, although Graham \etal (1990) 
suggested that debris disk may consist of $10-100$ cm particles.
If $a$ is several cm or larger (so that $R^\prime\gtrsim 1$) 
then $\psi_{Y,max}\sim 0.1$ for fully illuminated 
particles at $R_s$, see Fig. 3 of Vokrouhlick\'y (1998). 

However,
in the dense debris disk only the upper
(or lower) parts of particles, close to their spin axes, are 
directly illuminated by anisotropic starlight. This is because 
the latter illuminates the disk at small angle $\alpha$ and 
most of particle surface is shadowed by other particles 
nearby (thermal emission 
of these particles is on average isotropic in horizontal 
direction). From heuristic arguments one expects stellar 
heating to be mainly deposited at small 
colatitude $\sim \alpha^{1/2}\ll 1$ from the particle spin 
axis. It is easier for thermal conduction to isotropize 
surface temperature distribution around the spin axis over 
this small polar cap region rather than over the whole particle 
surface (as is assumed in standard calculation of $\psi_Y$), 
and this additionally lowers $\psi_Y$.

Thus, even if material properties of disk particles are favorable
for maximizing $\psi_Y$ (as is the case for our estimates here)
one still should expect $\psi_Y\lesssim 0.01$. Equations 
(\ref{eq:alpha}) and (\ref{eq:Mdot_r}) result in the 
following expression for Yarkovsky-induced outward mass 
accretion rate at $R_s$:
\ba
%
|\dot M_{Y,s}|&=&\frac{2^{9/2}\phi_r}{3}\psi_Y
\frac{\sigma R_\star^{5/2} T_\star^3 T_s}{c(GM_\star)^{1/2}}
\label{eq:massfluxYs}\\
&\approx &2.4\times 10^8\mbox{g s}^{-1}\phi_r\frac{\psi_Y}{0.01}
\frac{R_{\star,-2}^{5/2}T_{\star,4}^{3}}{M_{\star,1}^{1/2}}
\frac{T_s}{1500\mbox{K}}.
\nonumber
\ea
This is somewhat higher than $\dot M_{PR,s}$, thanks to the
high adopted $\psi_Y=10^{-2}$ exceeding $\psi_{PR}$, 
see (\ref{eq:psiPR}). However, this estimate 
may be very optimistic because of our poor knowledge of 
material properties and sizes of disk particles. 

Finally, we note that observations do not reveal the existence 
of {\it metal-poor} WDs with debris disks around them, which 
could plausibly exist if outward migration of particles due 
to the Yarkovsky force were effective at stopping the PR-driven 
inward mass accretion. This provides additional evidence that 
Yarkovsky effect does not play significant role in the 
circum-WD debris disk evolution.


\section{Comparison with observations.}
\label{sect:observations}

\begin{figure}
\plotone{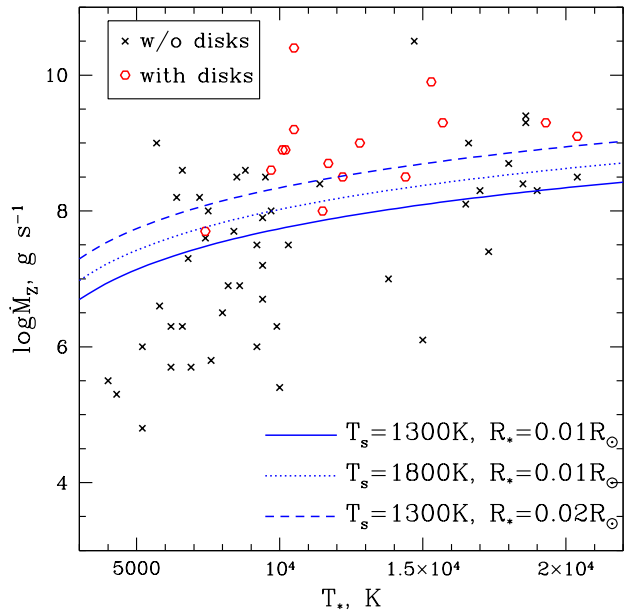}
\caption{
Mass accretion rates of high-Z elements $\dot M_Z$ 
inferred from elemental abundances measured in WD 
atmospheres. Metal-rich WDs without debris disks around 
them (with no detected IR excess) are depicted as crosses,
WDs with disks are displayed by open points. Curves show 
analytical prediction for $\dot M_{PR,s}$ (for different 
WD radii $R_\star$ and sublimation temperatures $T_s$)
given by equation (\ref{eq:massflux}). Note 
that all WDs with disks tend to lie above $\dot M_{PR,s}$
curve, implying that whenever a debris disk is present 
the Pointing-Robertson drag sets a lower limit on the metal 
accretion rate.
\label{fig:f2}}
\end{figure}

We now compare our theoretical predictions with data on 
$\dot M_Z$ inferred from observations\footnote{These estimates 
are somewhat model-dependent, see Jura \etal (2009b).} 
of metal-rich WDs. In Figure 
\ref{fig:f2} we plot the values of $\dot M_Z$ from Farihi \etal 
(2009, 2010) vs. stellar effective temperature $T_\star$ 
separately for WDs with and 
without debris disks detected via IR excesses. We also plot 
our analytical prediction for $\dot M_{PR,s}(T_\star)$ for 
different values of $R_\star$ and $T_s$. 

It is easy to see from Figure \ref{fig:f2} that all systems 
with detected debris disks exhibit $\dot M_Z$ higher than 
$\dot M_{PR,s}$. This is very encouraging since whenever WD 
is orbited by particulate disk one expects PR drag to set
a lower limit of the mass flux at the level 
of $\dot M_{PR,s}$. This lower bound on $\dot M_Z$ 
is hard (if at all possible) to avoid. This constraint  
naturally explains a strong positive correlation between 
$\dot M_Z$ and the rate of disk occurrence 
found by Farihi \etal (2010): the fraction of stars with 
compact debris disks is significantly higher for WDs with 
high $\dot M_Z$ simply because systems with disks cannot have
low $\dot M_Z<\dot M_{PR,s}$. All this strongly suggests that 
the PR drag indeed plays important and visible role in 
circum-WD debris disk evolution. 

It is worth emphasizing that the lack of disk-hosting systems 
at low $\dot M_Z$ is not caused by some  
observational bias: disk-bearing WDs with $\dot M_Z\sim 
\dot M_{PR,s}$ such as GD 16 (Farihi \etal 2009),
GD 56 (Jura \etal 2009a), G166-58 (Farihi \etal 2008) 
have (sometimes highly) significant detections of IR excesses by 
{\it Spitzer}.

It is also clear from Figure \ref{fig:f2} that there are metal-rich 
systems both with and without disks, in which $\dot M_Z$ significantly 
exceeds $\dot M_{PR,s}$ for a given $T_\star$, sometimes by several 
orders of magnitude. The most dramatic case is that of 
DAZB WD GD 362 which has $\dot M_Z\approx 2.5\times 10^{10}$ 
g s$^{-1}$ (Farihi \etal 2009),
exceeding $\dot M_{PR,s}$ corresponding to its $T_\star=10,500$ K 
by a factor of $\gtrsim 300$. 
Existence of such objects clearly implies that at least in some 
cases an additional accretion mechanism must operate on top 
of the PR drag, giving rise to very high   
$\dot M_Z$. We suggest such a mechanism that naturally operates in  
presence of a debris disk in Rafikov (2011, in preparation).

Systems without detected debris disks tend to occupy a broad range 
of $\dot M_Z$, both above and below $\dot M_{PR,s}$. They may
be interpreted as WDs that were orbited by compact debris disks 
in recent past but have now completely (or largely) lost their 
disks to accretion. Without active external source metals sediment 
out from atmospheres of these WD resulting in lower inferred 
$\dot M_Z$ in systems that had longer time since their debris disk 
disappearance.

Lifetime of a debris disk of mass $M_d$ accreting via
PR drag alone can be estimated as $M_d/\dot M_{PR}
\approx 2-4$ Myr for $M_d=10^{22}$ g (roughly corresponding to the 
mass of 200 km asteroid) and $\dot M_{PR}$ typical for a WD with 
$T_\star=10^4$ K. This estimate only marginally agrees with 
observational constraints suggesting several $10^5$ yr disk
lifetime (Farihi \etal 2009). However, one should keep in mind 
that this is an upper limit on the lifetime that is
realized if no other processes apart from the PR drag drive 
accretion of solids through the disk.


\section{Discussion.}
\label{sect:disc}

In conclusion we would like to emphasize the robust nature of the 
mass accretion due to PR drag. Our estimate (\ref{eq:massflux}) 
of $\dot M_{PR,s}$ depends neither on the disk properties 
such as $\Sigma_d$ (as long as $\tau_\parallel\gtrsim 1$) or 
$M_d$ nor on the material properties of its constituent 
particles. Also, 
$\dot M_{PR,s}$ does not vary much as $T_s$ or $R_\star$ 
change within reasonable limits.

One may worry that our calculation essentially disregards the 
details of the particle sublimation process. 
However, $\dot M_{PR,s}$ is set by stellar illumination 
{\it outside} the innermost annulus of directly exposed 
particles at the inner edge of the disk, see 
\S \ref{subsect:PR}. There sublimation is irrelevant and 
our calculations are robust. Interior of that 
point $\dot M_Z$ does not change by continuity and, 
as a result, mass accretion rate ends up being insensitive 
to how exactly directly exposed particles sublimate. 

\acknowledgements

Author thanks Konstantin Bochkarev for useful discussions. 
The financial support of this work is provided by the Sloan 
Foundation and NASA via grant NNX08AH87G.




\end{document}